\documentclass[twocolumn]{aastex631}
\usepackage{enumerate}
\usepackage{amsmath,amsthm,amsfonts,amssymb,amscd}
\usepackage{threeparttable,booktabs}
\received{XXX XX, XXXX}
\revised{XXX XX, XXXX}
\accepted{XXX XX, XXXX}
\submitjournal{ApJ}

\shorttitle{PAC. III. Accurate measurement of Galaxy Stellar Mass Function}
\shortauthors{Xu Jing \& Gao}

\begin{document}
\defcitealias{2022ApJ...925...31X}{Paper I}
\title{Photometric Objects Around Cosmic Webs (PAC) Delineated in a Spectroscopic Survey. III. Accurate Measurement of Galaxy Stellar Mass Function with the Aid of Cosmological Redshift Surveys}

\correspondingauthor{Y.P. Jing}
\email{ypjing@sjtu.edu.cn}

\author[0000-0002-7697-3306]{Kun Xu}
\affil{Department of Astronomy, School of Physics and Astronomy, Shanghai Jiao Tong University, Shanghai, 200240, People’s Republic of China}

\author[0000-0002-4534-3125]{Y.P. Jing}
\affil{Department of Astronomy, School of Physics and Astronomy, Shanghai Jiao Tong University, Shanghai, 200240, People’s Republic of China}
\affil{Tsung-Dao Lee Institute, and Shanghai Key Laboratory for Particle Physics and Cosmology, Shanghai Jiao Tong University, Shanghai, 200240, People’s Republic of China}
\author{Hongyu Gao}
\affil{Department of Astronomy, School of Physics and Astronomy, Shanghai Jiao Tong University, Shanghai, 200240, People’s Republic of China}



\begin{abstract}
We present a novel method to accurately measure the galaxy stellar mass function (GSMF) based upon the Photometric objects Around Cosmic webs (PAC) method developed in our first paper (Paper I) of the series. The method allows us to measure the GSMF to a lower mass end that is not accessible to the spectroscopic sample used in the PAC. Compared with Paper I, the current measurement of GSMF is direct and model independent. We measure the GSMFs in the redshift ranges of $z_s$\footnote{Throughout the paper, we use $z_s$ for spectroscopic redshift, $z$ for the z-band magnitude.}$<0.2$, $0.2<z_s<0.4$ and $0.5<z_s<0.7$ down to the stellar mass $M_*=10^{8.2}$, $10^{10.6}$ and $10^{10.6}M_{\odot}$, using the data from the DESI Legacy Imaging Surveys and the spectroscopic samples of Slogan Digital Sky Survey (i.e. Main, LOWZ and CMASS samples). Our results show that there is no evolution of GSMF from $z_s=0.6$ to $z_s=0.1 $ for $M_*>10^{10.6} M_{\odot}$, and that there is a clear up-turn at $M_*\approx 10^{9.5} M_{\odot}$ towards smaller galaxies in the local GMSF at $z_s=0.1$.  We provide an accurate double Schechter fit to the local GSMF for the entire range of  $M_*$ and a table of our measurements at the three redshifts, which can used to test theories of galaxy formation. Our method can achieve an accurate measurement of GSMF to the stellar mass limit where the spectroscopic sample is already highly incomplete (e.g. $\sim 10^{-3}$) for its target selection.  
\end{abstract}

\section{Introduction} \label{sec:intro}
Galaxy stellar mass function (GSMF) is one of the most crucial measurements for our understanding of galaxy evolution. The amplitude and shape of the GSMF encode vital information about the galaxy star formation and quenching mechanisms, and have been used to infer and constrain the galaxy formation models in many studies. 

According to the shapes of the GSMFs of star-forming and quenched galaxies, \citet{2010ApJ...721..193P} provided a model that separates the mass and environment quenching of galaxies at the local universe. Combined with the cosmic star formation rate, \citet{2013ApJ...770...57B} used the GSMFs at $z_s=0\sim8$ to constrain the star formation histories as a function of halo mass. \citet{2013MNRAS.428.3121M} determined the evolution of the galaxy-halo mass relation (SHMR) with the GSMFs at $z_s=0\sim4$. In many studies \citep{2006MNRAS.371..537W,2010MNRAS.402.1796W,2012ApJ...752...41Y,2022ApJ...925...31X,2022ApJ...928...10G}, the GSMF is used to accurately model the galaxy-halo connection combined with the galaxy correlation functions. Moreover, hydrodynamic simulations \citep{2015MNRAS.446..521S,2018MNRAS.473.4077P,2019MNRAS.486.2827D} and semi-analytic models \citep{2008MNRAS.391..481S,2012NewA...17..175B,2020MNRAS.491.5795H} are often calibrated and tested by the GSMF.

Therefore, accurately measuring the GSMFs at different redshifts is necessary for galaxy formation and evolution studies. In the past two decades, an enormous amount of efforts have been invested to measure the GSMF using different surveys and methods. 

At low redshift ($z_s<0.1$), large spectroscopic surveys, in particular the Two Degree Field Galaxy spectroscopic survey (2dF-GRS, \citealt{2001MNRAS.328.1039C}), Slogan Digital Sky Survey (SDSS, \citealt{2000AJ....120.1579Y}) and Galaxy and Mass Assembly (GAMA, \citealt{2011MNRAS.413..971D}), are used to study the local GSMF down to $10^{8.0}M_{\odot}$ \citep{2001MNRAS.326..255C,2003ApJS..149..289B,2003ApJ...592..819B,2008MNRAS.388..945B,2009MNRAS.398.2177L,2012MNRAS.421..621B,2016MNRAS.459.2150W,2022MNRAS.513..439D}. The GSMF is found to be well characterized by a double Schechter function. However, although the survey areas are large, the survey volumes are still limited due to the shallow survey depths. For example,  as will be demonstrated in Figure 3 below, the low mass galaxies ($<10^{9.0}M_{\odot}$) are only complete to $z_s \approx 0.03$ in the SDSS Main sample. The massive galaxies are also very few for their small number density. Thus, there are  large uncertainties in the GSMF in both the small and the large mass ends.  

At intermediate redshifts ($z_s<1$), deeper spectroscopic surveys begin to dominate the GSMF studies \citep{2007A&A...474..443P,2007ApJ...665..265F,2009ApJ...707.1595D,2010A&A...523A..13P,2013A&A...558A..23D,2013ApJ...767...50M}, such as the DEEP2
Galaxy spectroscopic survey \citep{2003SPIE.4834..161D}, the VIMOS VLT Deep Survey (VVDS, \citealt{2005A&A...439..845L}), zCOSMOS \citep{2007ApJS..172...70L}, The PRIsm MUlti-object Survey (PRIMUS, \citealt{2011ApJ...741....8C}) and the VIMOS Public Extragalactic spectroscopic survey (VIPERS, \citealt{2014A&A...562A..23G}). The GSMF has been explored down to $10^{10.0}M_{\odot}$ using these surveys. However, despite the huge efforts, measuring low mass objects is still very difficult, and stellar mass limited samples are usually very small. In addition, incompleteness should be considered with caution when measuring the GSMF with the spectroscopic samples. A few effects such as target selection, fiber collision, target sampling rate and spectroscopic success rate should be carefully considered, and the derived GSMF is sensitive to the details of the corrections for the incompleteness. Large spectroscopic surveys, such as Baryon Oscillation Spectroscopic Survey (BOSS, \citealt{2015ApJS..219...12A,2016MNRAS.455.1553R}), are also used to study the GSMF at the massive end at intermediate redshifts \citep{2013MNRAS.435.2764M}. Due to the effects mentioned above, \citet{2016MNRAS.457.4021L} found that CMASS and LOWZ samples are still incomplete for massive galaxies ($>10^{11.3}M_{\odot}$). To derive the GSMF from BOSS, \citet{2018ApJ...858...30G} modeled the galaxy clustering and incompleteness simultaneously using the incomplete conditional stellar mass function (ICSMF) framework, and inferred the GSMF from the constrained SHMR. This method attempted to recover the incompleteness, but their result is model dependent.

Deep multi-band photometric surveys, which are deeper and more complete in terms of galaxy populations, are also used to study the GSMF at intermediate and high redshifts. Combining the ground-based deep photometric surveys and Hubble Space Telescope (HST) near-IR imaging, many studies push the measurements of the GSMF to $z_s>1$ \citep{2006A&A...459..745F,2013A&A...556A..55I,2013ApJ...777...18M,2014ApJ...783...85T,2015MNRAS.447....2M,2017A&A...605A..70D,2018MNRAS.480.3491W,2020ApJ...893..111L,2021MNRAS.503.4413M,2022arXiv220310895S}. The survey areas of the deep multi-band photometric surveys are even smaller ($1\sim2\ \rm{deg}^2$) than the deep spectroscopic surveys, since they require long exposure times to reach very faint sources and multiple bands from UV to IR to obtain relatively accurate photometric redshifts (photo-z) for faint objects. Wide photometric surveys are usually not suitable for the GSMF studies due to the fewer observed bands and shallower depths. Therefore, the cosmic variance should be taken carefully into account when measuring the GSMF using deep photometric surveys, especially at the high mass end. Moreover, although faint sources can be detected in photometric surveys, photometric redshifts used in the GSMF studies are usually trained with the spectroscopic data, which may not cover for very faint sources, and therefore have larger errors. These problems associated with the photometric redshifts will also introduce uncertainties in the GSMF.   

Furthermore, studying the evolution of the GSMF using the measurements from different surveys \citep{2013ApJ...770...57B,2013MNRAS.428.3121M} should be very cautious especially at the high mass end, where the GSMF changes exponentially. Small systematics, from such as photometric calibration, source extraction and modeling, incompleteness correction and stellar mass estimation method, can cause large difference in the GSMF.

In \citet[hearafter \citetalias{2022ApJ...925...31X}]{2022ApJ...925...31X}, on the basis of \citet{2011ApJ...734...88W}, we developed a method named Photometric objects Around Cosmic webs (PAC) to estimate the excess surface density $\bar{n}_2w_p$ of photometric objects with certain physical properties around spectroscopically identified sources, which can take full use of the spectroscopically and deeper photometric surveys. In Paper I and the second paper \citep{2022ApJ...926..130X}, based on PAC measurements, we have studied the SHMR and the galaxy assembly bias of massive galaxies. Based on the SHMR, we predicted the GSMF which is in good agreement with the observations from the literature. In this third paper, we will combine the PAC measurements of $\bar{n}_2w_{\rm{p}}$ and the projected cross-correlation function $w_{\rm{p}}$ measured from the spectroscopic samples, and derive $\bar{n}_2$, the GSMF in the photometric survey. The measurements of $w_{\rm{p}}$ are insensitive to the incompleteness of the spectroscopic samples in the galaxy stellar mass, once the sample can be used to accurately measure the projected cross-correlation function $w_{\rm{p}}$ (which is usually be the case for cosmological redshift surveys). With this method, we can extend the study of GSMF to lower mass end with the wide spectroscopic and deeper photometric surveys. We can also provide  measurements of the GSMFs in a uniform way (e.g. the same photometric catalog, the same stellar mass method) to redshift $z_s=0.6$  which is better for evolution studies. Compared with the model prediction of GSMF in Paper I, the current measurement of GSMF is direct and model independent.

We introduce our method in Section \ref{sec:method}. In Section \ref{sec:data}, we describe the data and designs used to measure the GSMF. The results are shown in Section \ref{sec:results}, and our conclusion is given in Section \ref{sec:conclusion}. We adopt the cosmology with $\Omega_m = 0.268$, $\Omega_{\Lambda} = 0.732$ and $H_0 = 71{\rm \ km/s/Mpc}$ throughout the paper.

\section{Methodology} \label{sec:method}
\subsection{Photometric objects Around Cosmic webs (PAC)}
Within a relatively narrow redshift range, supposing we want to study two types of galaxies with $\rm{pop}_1$ from a spectroscopic catalog and $\rm{pop}_2$ from a deep photometric catalog. In \citetalias{2022ApJ...925...31X}, we provide a method called PAC that can measure the excess projected density distribution $\bar{n}_2w_{\rm{p}}(r_{\rm{p}})$ of $\rm{pop}_2$ with certain physical properties around $\rm{pop}_1$:
\begin{equation}
    \bar{n}_2w_{\rm{p}}(r_{\rm{p}}) = \frac{\bar{S}_2}{r_1^2}w_{12,\rm{weight}}(\theta)\,\,,\label{eq:1}
\end{equation}
where $r_1$ is the comoving distance to $\rm{pop}_1$, $w_{\rm{p}}(r_{\rm{p}})$ and $w_{12,\rm{weight}}(\theta)$ are the projected cross-correlation function (PCCF) and the weighted angular cross-correlation function (ACCF) between $\rm{pop}_1$ and $\rm{pop}_2$ with $r_{\rm{p}}=r_1\theta$, and $\bar{n}_2$ and $\bar{S}_2$ are the mean number density and mean angular surface density of $\rm{pop}_2$. Since $\rm{pop}_1$ has a redshift distribution, $w_{12}(\theta)$ is weighted by $1/r^2_1$ to account for the effect that $r_{\rm{p}}$ varies with redshift at fixed $\theta$. The advantage of PAC is to estimate the rest-frame physical properties of $\rm{pop}_2$ statistically without the need of photo-z, so that we can take full use of the deep photometric surveys. The main steps of PAC are:
\begin{enumerate}[(i)]
    \item Split $\rm{pop}_1$ into narrower redshift bins, mainly accounting for the fast change of $r_1$ with redshift.
    \item Assuming all galaxies in $\rm{pop}_2$ have the same redshift as the mean redshift in each redshift bin, calculate the physical properties of $\rm{pop}_2$ using methods such as spectral energy distribution (SED). Thus, in each redshift bin of $\rm{pop}_1$, there is a physical property catalog of $\rm{pop}_2$ .
    \item In each redshift bin, select $\rm{pop}_2$ with certain physical properties and calculting $\bar{n}_2w_{\rm{p}}(r_{\rm{p}})$ according to Equation \ref{eq:1}. The foreground and background objects with wrong properties are cancelled out through ACCF and only $\rm{pop}_2$ around $\rm{pop}_1$ with correct redshifts left.
    \item Combine the results from different redshift bins by averaging with proper weights.
\end{enumerate}
For more details, we refer to \citetalias{2022ApJ...925...31X}.

\subsection{Estimating the galaxy stellar mass function}
Using PAC, we can calculate $\bar{n}_2w_{\rm{p}}(r_{\rm{p}})$ for $\rm{pop}_2$ with certain stellar mass. And we can also select a sample from the spectroscopic catalog with the same stellar mass. To distinguish these two samples, in the following, we denote the samples from the photometric catalog and the spectroscopic catalog as $\rm{pop}_2^p$ and $\rm{pop}_2^s$ respectively. We can calculate the PCCF $w_{\rm{p}}(r_{\rm{p}})$ between $\rm{pop}_1$ and $\rm{pop}_2^s$, and derive the number density $\bar{n}_2$ by comparing it to the PAC measurement $\bar{n}_2w_{\rm{p}}$.

The GSMF can be obtained by deriving $\bar{n}_2$ for $\rm{pop}_2$ with different stellar masses. The low mass end of the GSMF is still limited by the spectroscopic catalog, since the number of galaxies at the low mass end may not even be enough to get $w_{\rm{p}}(r_{\rm{p}})$.

With $\rm{pop}_1$ and $\rm{pop}_2^s$ from spectroscopic catalogs at different redshifts and a deep photometric catalog, we can also measure the time evolution of the GSMF using PAC. 

The whole process is straightforward and model independent with the only assumption that $\rm{pop}_2^s$ and $\rm{pop}_2^p$ have the same bias on large scales. According to the observed dependence \citep{2006MNRAS.368...21L} of galaxy clustering on the color and the modeling predictions \citep{2015ApJ...799..130R,2016MNRAS.457.4360Z}, we find that the linear bias of blue (red) population is smaller (or larger) than the total population  for $10^8  <M_*<10^{11} M_\odot$ by $<0.1dex$, with the largest difference found at the smallest stellar mass. Therefore, our above assumption may lead an overestimate (underestimate) of the GSMF by the same amount if the population in the spectroscopic $\rm{pop}_2^s$ is dominated by blue (red) ones. This amount of systematics can be tolerated in the precision of our current measurement (cf Figure 4). With the accuracy expected to improve in future surveys, the linear clustering bias of $\rm{pop}_2^s$ relative to  $\rm{pop}_2^p$ should be carefully modeled. 

\section{Data and designs} \label{sec:data}
In this section, we introduce the observational data used in this work, and the details of the PAC and $w_{\rm{p}}$ measurements. Systematic errors and incompleteness of photometric and spectroscopic data are also carefully considered.
\subsection{Photometric data}
We use the photometric catalogs\footnote{https://www.legacysurvey.org/dr9/catalogs/} from the DR9 of DESI Legacy Imaging Surveys \citep{2019AJ....157..168D} throughout the paper. They cover over $14000\ \rm{deg}^2$ of the sky in the Dark Energy Spectroscopic Instrument (DESI; \citealt{2016arXiv161100036D}) footprint and consist of three different components:
\begin{enumerate}[(i)]
    \item The Dark Energy Camera Legacy Survey (DECaLS), which observes around $9000\ \rm{deg}^2$ in both the Northern and Southern Galactic caps (NGC and SGC) at $\rm{Dec}\leq32\ \rm{deg}$ in $g$, $r$ and $z$ band. It also includes the data from the deeper Dark Energy Survey (DES; \citealt{2016MNRAS.460.1270D}) in the SGC covering around $5000\ \rm{deg}^2$, where $1130\ \rm{deg}^2$ is in the DESI footprint.
    \item The Beijing-Arizona Sky Survey (BASS), which covers
    $5000\ \rm{deg}^2$ in the NGC footprint at $\rm{Dec}\geq32\ \rm{deg}$ in $g$ and $r$ band but is around $0.5\ \rm{mag}$ shallower than DECaLS. It also observes additional $500\ \rm{deg}^2$ in the DECaLS footprint in order to understand and correct the systematic biases.
    \item The Mayall z-band Legacy Survey (MzLS), which observes the same footprint as BASS in the NGC footprint at $\rm{Dec} \geq 32\ \rm{deg}$ in the z band. The $z$ band depth is comparable to DECaLS.
\end{enumerate}

Altogether, they provide three band ($g$, $r$ and $z$) photometric data over more than $14000\ \rm{deg}^2$ (with additional $5000\ \rm{deg}^2$ from DES) with a $5\sigma$ point source depth of $g=24.9$, $r=24.2$ and $z=23.3$ for DECaLS and MzLS, but BASS is around $0.5\ \rm{mag}$ shallower. 

\begin{figure}
    \plotone{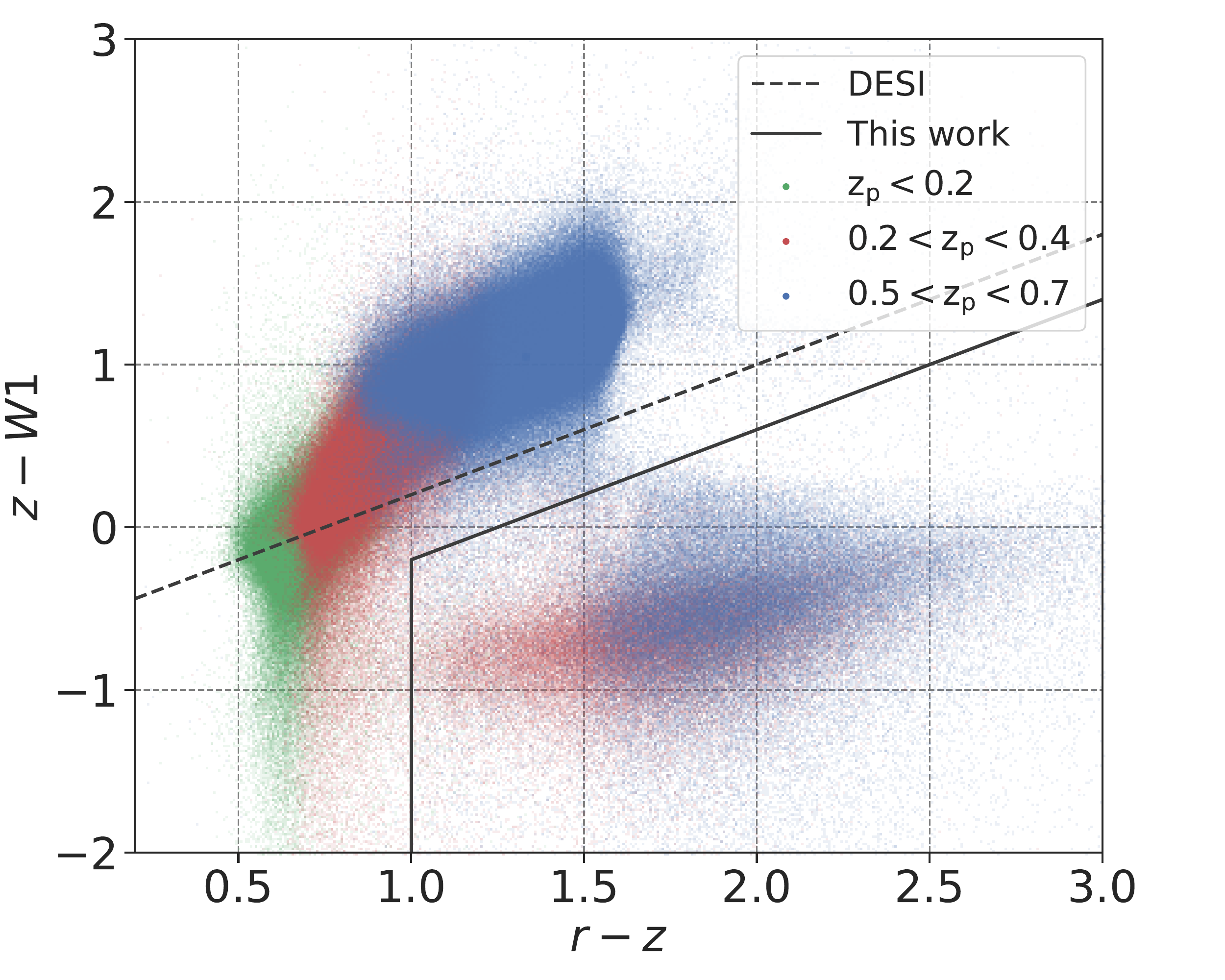}
    \caption{The $r-z$ vs. $z-W1$ color-color diagram for sources with $z<22.0$ from DECaLS color-coded according to their photometric redshift \citep{2021MNRAS.501.3309Z}. Dashed line is the non-stellar cut adopted in DESI LRG target selection \citep{2020RNAAS...4..181Z} and solid lines are the cuts adopted in this work.}
    \label{fig:star}
\end{figure}

The Legacy Surveys images are processed using \texttt{Tractor}\footnote{https://github.com/dstndstn/tractor} \citep{2016ascl.soft04008L}, a forward-modeling approach to perform source extraction on pixel-level data. The sources are modeled with parametric profiles convolved with a specific point spread function (PSF), including a delta function for the point source, exponential law, de Vaucouleurs law,  and a S\'ersic profile. And the sources are classified into six morphological types according to the best-fit models: point sources (PSF), round exponential galaxies (REX), de Vaucouleurs profiles (DEV), normal exponential profiles (EXP), and S\'ersic profiles (SER).

\begin{figure*}
    \plottwo{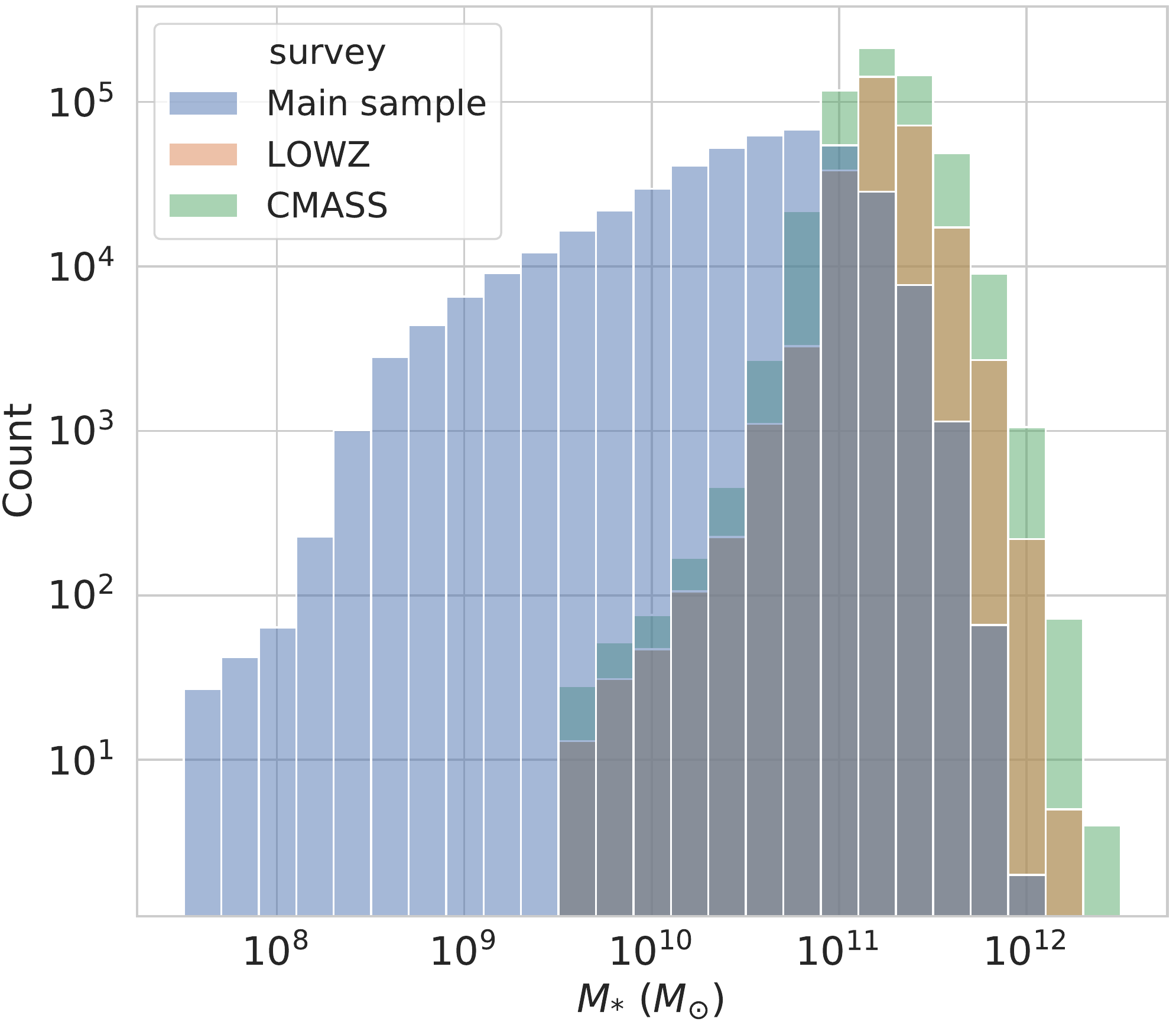}{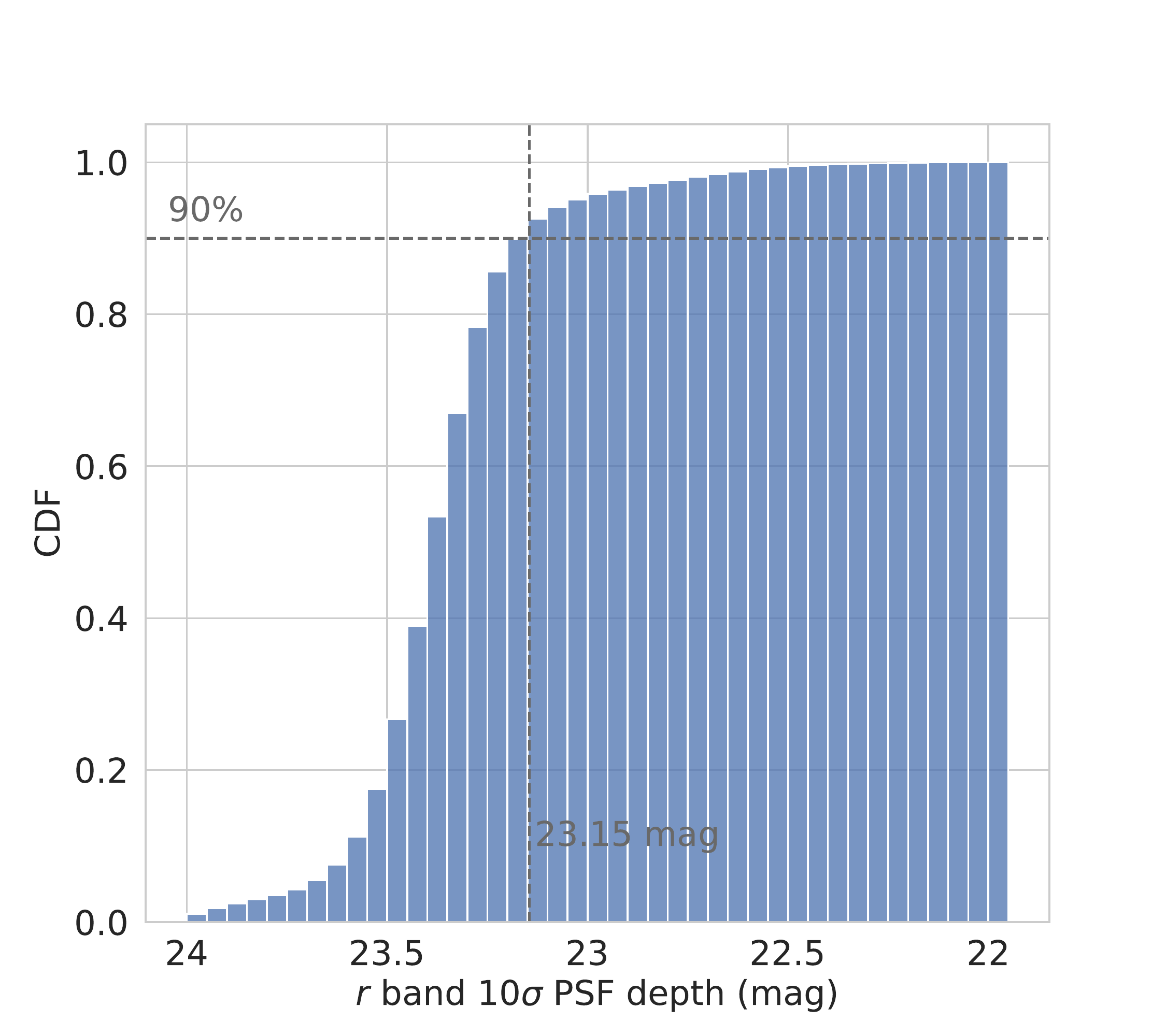}
    \caption{Left: Stellar mass distributions for the Main, LOWZ and CMASS spectroscopic samples. Right: Cumulative distribution function of the $r$ band $10\sigma$ PSF depth for DECaLS.}
    \label{fig:fig2}
\end{figure*}

We only use the footprints that have been observed at least once in all three bands, and perform bright star mask and bad pixel mask to the catalogs using the MASKBITS\footnote{https://www.legacysurvey.org/dr9/bitmasks/} provided by the Legacy Surveys. The photometric catalog is further selected to match the survey geometry of the spectroscopic data at each redshift. Galactic extinction is corrected for all the sources using the maps of \citet{1998ApJ...500..525S}. 

To reject stars, we exclude sources with PSF morphologies. However, while it can remove most of the stars, there are still some stellar objects with extended morphologies left, such as close binaries. Therefore, following DESI LRG target selection \citep{2020RNAAS...4..181Z}, with $W1$ band data from WISE \citep{2010AJ....140.1868W}, we adopt color cuts in $r-z$ vs. $z-W1$ diagram to further remove the stellar objects. We adopt color cuts different from \citet{2020RNAAS...4..181Z} to preserve more galaxies especially at low $z_s$. The color cuts are shown in Figure \ref{fig:star} with stars have
\begin{equation}
    (z-W1<0.8\times(r-z)-1.0)\ {\rm{AND}}\ (r-z>1.0)\,\,.
\end{equation}
We would like to point out that since stars are not correlated with spectroscopic galaxies, our PAC results are insensitive to the color cuts. But these cuts help to remove the stars when we use the photometric catalog with photo-z in \S4.2.

\subsection{Spectroscopic data}
Spectroscopic catalogs used for PAC in this work are all from SDSS \citep{2000AJ....120.1579Y}). We focus on three redshift bins $0<z_s<0.2$, $0.2<z_s<0.4$ and $0.5<z_s<0.7$ using data from SDSS DR7 Main sample \citep{2009ApJS..182..543A} and SDSS-III BOSS DR12 LOWZ and CMASS samples \citep{2015ApJS..219...12A,2016MNRAS.455.1553R}, respectively.

The DR7 main sample provides a spectroscopic galaxy catalog complete to a {\tt{Petrosian}} \citep{1976ApJ...209L...1P} magnitude limit of $r = 17.77$, which covers over $7500\ \rm{deg}^2$ of the NGC. It also includes three stripes covering additional $532\ \rm{deg}^2$ in the SGC, which are not considered in this work. We use the "{\tt{bright}}" LSS catalog\footnote{http://sdss.physics.nyu.edu/lss/dr72/bright/} of the Main sample with collision corrections and a constant flux limit of $r<17.6$. We adopt a redshift cut of $z_s<0.2$ for the Main sample.

The LOWZ sample is designed to extend the SDSS-I/II Cut I LRG
sample \citep{2001AJ....122.2267E} to $z_s \approx 0.4$ and to fainter luminosities. It adopts a color-magnitude cut on galaxies with {\tt{cmodel}} \citep{2004AJ....128..502A} brightness limits of $16<r<19.6$. We refer to \citet{2016MNRAS.455.1553R} for the full target selection strategy. We use the "{\tt{LOWZ}}" LSS catalog\footnote{https://data.sdss.org/sas/dr12/boss/lss/} in BOSS DR12 for the LOWZ sample. The galaxies from the first nine months of the BOSS observation are excluded in the "{\tt{LOWZ}}" catalog due to the incorrect star–galaxy separation criterion, resulting in a smaller footprint compared to "{\tt{CMASS}}". In total, it covers $8337\ \rm{deg}^2$ with $5836\ \rm{deg}^2$ in the NGC and $2501\ \rm{deg}^2$ in the SGC. We adopt a redshift cut of $0.2<z_s<0.4$ for the LOWZ sample.

The CMASS sample uses target selections similar to those of the SDSS-I/II Cut II LRG sample, but is bluer and fainter to increase the galaxy number
density in the redshift range of $0.4<z_s<0.7$. Galaxies in the CMASS are selected with a number of magnitude and colour cuts to get an approximately constant stellar mass. The {\tt{cmodel}} magnitude limits for CMASS sample are $17.6<i<19.9$ and the full selection criteria can be found in \citet{2016MNRAS.455.1553R}. We use the "{\tt{CMASS}}" LSS catalog in BOSS DR12 for the CMASS sample, which covers $9376\ \rm{deg}^2$ of the sky with $6851\ \rm{deg}^2$ in the NGC and $2525\ \rm{deg}^2$ in the SGC. We adopt a redshift cut of $0.5<z_s<0.7$ for the CMASS sample.

All the three spectroscopic samples are within the footprint of the DESI Legacy Imaging Surveys. The spectroscopic sources are matched with the photometric catalog described above to get $g,r\ \rm{and}\ z$ band flux measurements.

\subsection{Spectral Energy Distribution (SED)}
Physical properties of the spectroscopic and photometric sources are estimated using the SED code {\tt{CIGALE}} \citep{2019A&A...622A.103B} with the $grz$ band measurements.

We use the \citet{2003MNRAS.344.1000B} stellar population synthesis models with a \citet{2003PASP..115..763C} initial mass function and a delayed star formation history $\phi(t)\approx t\exp{-t/\tau}$. Three metallicities, $Z/Z_{\odot}=0.4,\ 1\ \rm{and}\ 2.5$, are considered, where $Z_{\odot}$ is the metallicity of the Sun. The \citet{2000ApJ...533..682C} extinction law with $0<E(B-V)<0.5$ is adopted for dust reddening. We use the {\tt{bayes}} type of outputs from the {\tt{CIGALE}} in this work.

\begin{figure*}[t]
    \plotone{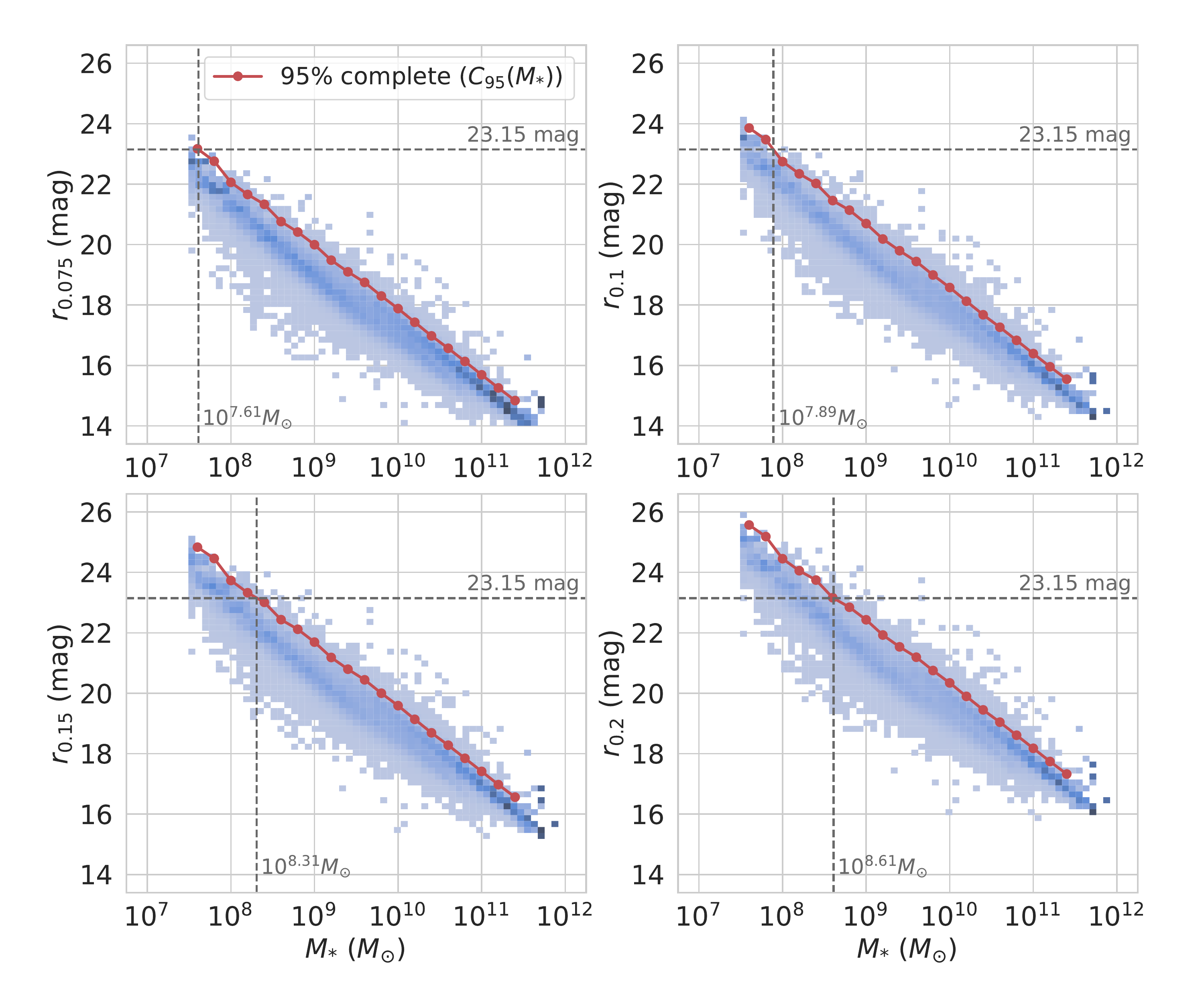}
    \caption{Stellar mass - $r$ band magnitude relations for GAMA DR4 galaxies k-corrected to $z_s=0.075,\ 0.1,\ 0.15\ {\rm{and}}\ 0.2$. Galaxies are weighted by $1/V_{\rm{max}}$. Red lines with dots show the $95\%$  completeness limits $C_{95}(M_{*})$ varying with the stellar mass at each redshifts. Grey dashed lines show the stellar mass limits for DECLaS with the $r$ band depth of $23.15\ \rm{mag}$.}
    \label{fig:fig3}
\end{figure*}

\subsection{Completeness and designs}
Despite the huge efforts of the survey teams, as shown in Appendix \ref{sec:C}, small but significant systematics still exist between the photometric measurements of DECaLS and BASS+MzLS, resulting in an offset in the stellar mass estimation. As shown in Appendix \ref{sec:C}, we use both DECaLS and BASS+MzLS for the $w_{\rm{p}}$ measurements and use only DECaLS for the PAC $\bar{n}_2w_{\rm{p}}$ measurements, because the offset has a much larger influence on $\bar{n}_2$ especially at the high mass end, due to the exponential change of the GSMF.

Based on the test in \citetalias[see Figure 3 there]{2022ApJ...925...31X}, we split both the LOWZ and CMASS samples into two narrower redshift bins with equal bin width for the PAC measurements. Since the comoving distance changes faster at the low redshifts, the Main sample is split into four redshift bins $[0.05,0.075],[0.75,0.1],[0.1,0.15]\ \rm{and}\ [0.15,0.2]$, and galaxies with $z_s<0.05$ are not used in the PAC measurements.

In the left panel of Figure \ref{fig:fig2}, we show the stellar mass distribution of the three spectroscopic samples. According to the number of galaxies, we choose $[10^{8.1},10^{11.7}]M_{\odot}$, $[10^{10.5},10^{11.9}]M_{\odot}$ and $[10^{10.5},10^{12.1}]M_{\odot}$ as the stellar mass ranges of $\rm{pop}_2$ for the Main sample, LOWZ and CMASS respectively, and split them into smaller mass bins with an equal logarithmic interval of 0.2 in $\log M_*$. They are also the stellar mass ranges we can calculate for the GSMFs at each redshift. 

DECaLS is deep enough to reach $10^{10.5}M_{\odot}$ at the redshifts of LOWZ and CMASS (\citetalias[see Figure 1]{2022ApJ...925...31X}). However, whether DECaLS is complete or not for $10^{8.1}M_{\odot}$ at $z_s<0.2$ remains to be explored. As in \citetalias{2022ApJ...925...31X}, we use the $10\sigma$ PSF depth as the depth for extended sources, which is more stringent than the {\tt{galdepth}} provided by DECaLS. In the right panel of Figure \ref{fig:fig2}, we show the Cumulative Distribution Function (CDF) of the $r$ band $10\sigma$ PSF depth for DECaLS. We find that $90\%$ of the regions in DECaLS are deeper than $23.15\ \rm{mag}$. Thus, we use $r=23.15$ as the galaxy depth for DECaLS.

Deep photometric samples with photo-z as used in \citetalias{2022ApJ...925...31X} are no longer suitable for completeness studies at $z_s<0.2$, since the survey volumes become too small and the photo-z results in larger errors in the distance estimate. Instead, we use the deeper Galaxy And Mass Assembly (GAMA) DR4 spectroscopic sample \citep{2022MNRAS.513..439D}, which covers around $250\ \rm{deg}^2$ of the sky with a $95\%$ completeness to $r_{\rm{KiDS}}=19.65$. The DR4 of GAMA replaced the previously used SDSS optical band ($ugri$) data with the much deeper Kilo Degree Survey (KiDS) DR4 data \citep{2019A&A...625A...2K,2020MNRAS.496.3235B}, providing more reliable photometric measurements for the faint sources.

We calculate the stellar mass for galaxies in GAMA using SED and output the best-fit spectra. The spectra are redshifted to $z_s=0.075,\ 0.1,\ 0.15\ \rm{and}\ 0.2$ and convolved with the DECaLS $r$ band filter to get the DECaLS $r$ band magnitudes, corresponding to the k-corrections \citep{1996ApJ...467...38K,2007AJ....133..734B}. We also calculate the $V_{\rm{max}}$ for each galaxy, where $V_{\rm{max}}$ is the volume corresponding to $z_{\rm{max}}$, the maximum redshift over which the galaxy can pass the GAMA selection criteria.

In Figure \ref{fig:fig3}, we show the stellar mass - k-corrected $r$ band magnitude relations at four redshifts $z_s=0.075,\ 0.1,\ 0.15\ \rm{and}\ 0.2$ for GAMA galaxies weight by $1/V_{\rm{max}}$. Following \citetalias{2022ApJ...925...31X}, we calculate the $r$-band completeness limit $C_{95}(M_{*})$ that $95\%$ of the galaxies are brighter than $C_{95}(M_{*})$ in the $r$-band for a given stellar mass $M_{*}$ (red lines). For DECaLS with the $r$ band galaxy depth of $23.15\ \rm{mag}$, the complete stellar masses are $10^{7.61}M_{\odot}$, $10^{7.89}M_{\odot}$, $10^{8.31}M_{\odot}$ and $10^{8.61}M_{\odot}$ at redshift $0.075$, $0.1$, $0.15$ and $0.2$ (grey dashed lines), respectively. Thus, for the Main sample, $\bar{n}_2w_{\rm{p}}$ is only calculated at $z_s<0.1$ and $z_s<0.15$ for $\rm{pop}_2$ with mass of $10^{8.2}M_{\odot}$ and $10^{8.4}M_{\odot}$, and the whole redshift range is used for $\rm{pop}_2$ larger than $10^{8.6}M_{\odot}$.

Then, we choose the stellar mass range of $\rm{pop}_1$ as $[10^{10.3},10^{11.3}]M_{\odot}$ for the Main sample and $[10^{11.3},10^{11.9}]M_{\odot}$ for LOWZ and CMASS. To check the systematics of the data and our method, we also split $\rm{pop}_1$ into several stellar mass bins with an equal logarithmic interval of $0.2$. However, when cross-correlating with $\rm{pop}_2$ of the smallest or largest mass bin, it is hard to calculate $w_{\rm{p}}$ if the $\rm{pop}_1$ sample is divided as stated previously. In these cases, the whole spectroscopic samples are used.  

The final designs for the measurements of $w_{\rm{p}}$ and $\bar{n}_2w_{\rm{p}}$ are summarized in Table \ref{tab:t1}.

\begin{table*}
    \centering
    \caption{Final designs for the $w_{\rm{p}}$ (DECLaS and BASS+MzLS) and PAC (DECaLS) measurements.}
    \begin{threeparttable}
    \setlength{\tabcolsep}{6mm}{
    \begin{tabular*}{\hsize}{ccccccc}
     \toprule
     redshift & Survey & $\rm{pop}_1$\tnote{a} & $\rm{pop}_2$\tnote{b} & PAC redshift bins \\
      &  & ($M_{\odot}$) & ($M_{\odot}$) &  \\
     \midrule
     $[0,0.2]$ & Main & $[10^{10.3},10^{11.3}]$ & $[10^{8.1},10^{11.7}]$ & $[0.05,0.075],[0.75,0.1],[0.1,0.15],[0.15,0.2]$\\
     $[0.2,0.4]$ & LOWZ & $[10^{11.3},10^{11.9}]$ & $[10^{10.5},10^{11.9}]$ & $[0.2,0.3],[0.3,0.4]$ \\
     $[0.5,0.7]$ & CMASS & $[10^{11.3},10^{11.9}]$ & $[10^{10.3},10^{12.1}]$ & $[0.5,0.6],[0.6,0.7]$ \\
     \bottomrule
    \end{tabular*} }
    \begin{tablenotes}
     \footnotesize
     \item[a] Stellar mass ranges of $\rm{pop}_1$ with an fiducial equal logarithmic bin width of $10^{0.2}M_{\odot}$.
     \item[b] Stellar mass ranges of $\rm{pop}_2$ with an fiducial equal logarithmic bin width of $10^{0.2}M_{\odot}$.
    \end{tablenotes}
    \end{threeparttable}
    \label{tab:t1}
\end{table*}

\begin{figure*}
    \plotone{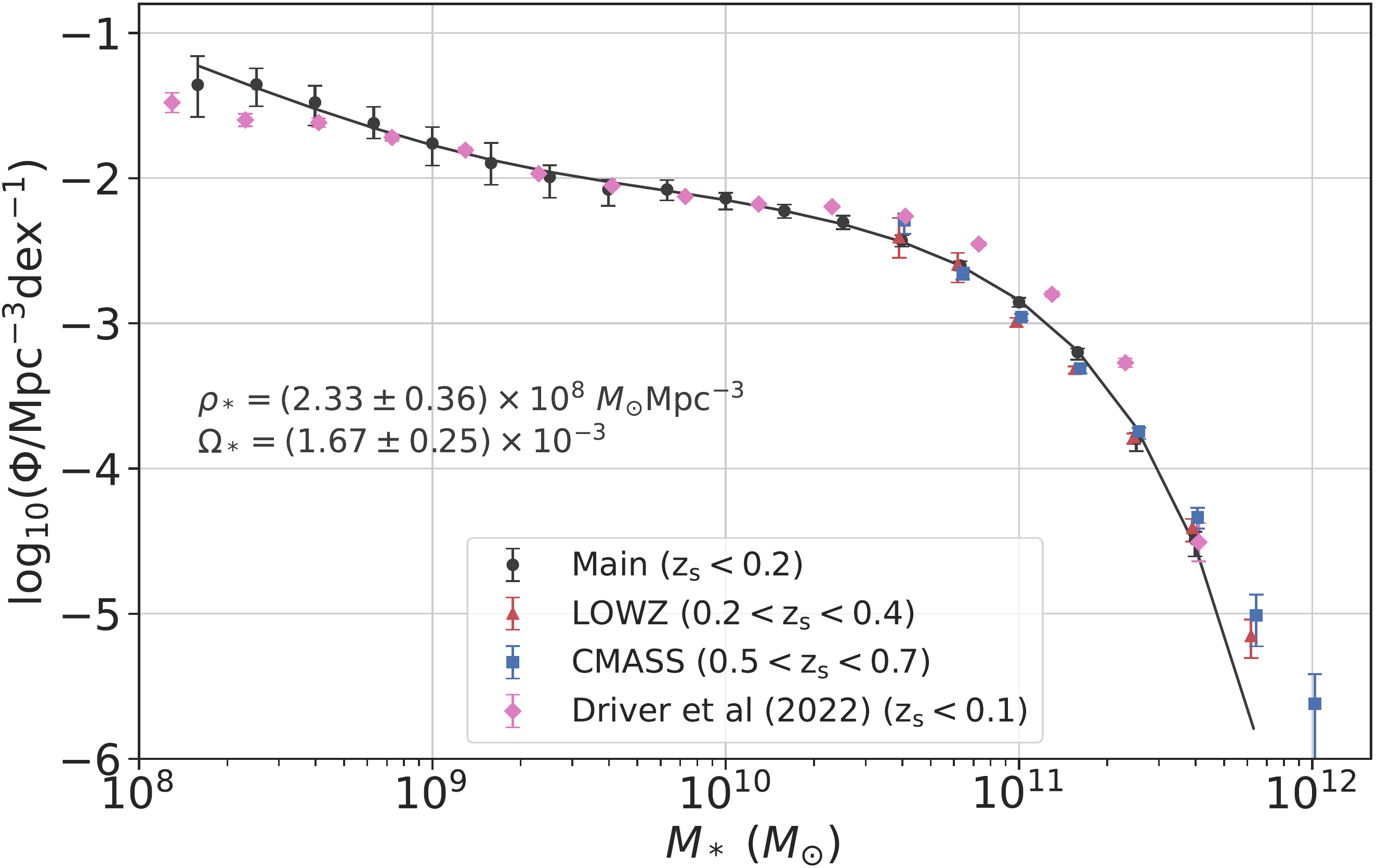}
    \caption{The GSMFs estimated according to the designs in Table \ref{tab:t1} using the Main sample ($z_s<0.2$, black), LOWZ ($0.2<z_s<0.4$, red) and CMASS ($0.5<z_s<0.7$, blue). Results from the Main sample are fitted by a double Schechter function shown in black line. The GAMA DR4 \citep{2022MNRAS.513..439D} measurements are provided for comparison with purple diamonds. The GSFD ($\rho_*$ and $\Omega_*$) from the Main sample results is also given.}
    \label{fig:fig4}
\end{figure*}

\subsection{Measurements of $\bar{n}_2w_{\rm{p}}$, $w_{\rm{p}}$ and $\bar{n}_2$}
For particular mass and redshift bin of $\rm{pop}_2^p$ and $\rm{pop}_1$, $\bar{n}_2w_{\rm{p}}$ is calculated in a few narrower redshift sub-bins for two separate regions (DECaLS NGC and DECaLS SGC), and then the results are combined. To properly account for the contributions from different redshift bins and from the two regions, we consider a more refined method than in \citetalias{2022ApJ...925...31X} to combine the $\bar{n}_2w_{\rm{p}}$ results. 
 
 Let $\mathcal{A}\equiv \bar{n}_2w_{\rm{p}}$ for better representation. Assuming $\mathcal{A}$ is calculated in $N_{\rm{r}}$ narrower redshift bins and $N_{\rm{s}}$ sky regions, we further split each sky regions into $N_{\rm{sub}}$ sub-regions for error estimation using jackknife resampling. $\mathcal{A}_{i,j,k}$ can be calculated according to Equation \ref{eq:1} in $i$th redshift bin, $j$th sky region and $k$th jackknife sub-samlpe. Measurements from different sky regions are first combined, weighted by the areas $w_{\rm{s}}$ of the regions: 
\begin{equation}
     \mathcal{A}_{i,k}=\frac{\sum_{j=1}^{N_{\rm{s}}}\mathcal{A}_{i,j,k}w_{\rm{s},j}}{\sum_{j=1}^{N_{\rm{s}}}w_{\rm{s},j}}\,\,.
\end{equation}
 Then, we estimate the mean values and the uncertainties of the mean values for each redshift bins from $N_{\rm{sub}}$ sub-samples:
\begin{equation}
     \mathcal{A}_i=\sum_{k=1}^{N_{\rm{sub}}}\mathcal{A}_{i,k}/N_{\rm{sub}}\,\,,
\end{equation}
\begin{equation}
    \sigma_{\mathcal{A},i}^2=\frac{N_{\rm{sub}}-1}{N_{\rm{sub}}}\sum_{k=1}^{N_{\rm{sub}}}(\mathcal{A}_{i,k}-\mathcal{A}_i)^2\,\,.
\end{equation}
Finally, results from different redshift bins are combined according to the uncertainties
\begin{equation}
    \mathcal{A}=\frac{\sum_{i=1}^{N_{\rm{r}}}\mathcal{A}_{i}/\sigma_{\mathcal{A},i}^2}{\sum_{i=1}^{N_{\rm{r}}}1/\sigma_{\mathcal{A},i}^2}\,\,,
\end{equation}
\begin{equation}
    \sigma_{\mathcal{A}}^2=\frac{1}{\sum_{i=1}^{N_{\rm{r}}}1/\sigma_{\mathcal{A},i}^2}\,\,.
\end{equation}

Using the spectroscopic samples $\rm{pop}_2^s$ and $\rm{pop}_1$, $w_{\rm{p}}$ and $\sigma_{w_{\rm{p}}}$ can be calculated in the same way. However, the samples are not splitted into narrower redshift bins as done in PAC, so the final step can be ignored or you may regard $N_{\rm{r}}=1$. $w_{\rm{p}}$ is estimated using the Landy–Szalay estimator \citep{1993ApJ...412...64L}.

$\mathcal{A}$ and $w_{\rm{p}}$ estimated above are only used for the tests of systematics and verification of modelings. We do not use the simple error propagation to calculate the $\bar{n}_2$ and its uncertainty. Instead, we use a more sophisticated way. After we combine the measurements from different sky regions and obtain $\mathcal{A}_{i,k}$ and $w_{{\rm{p}},k}$, we can calculate $\mathbf{\bar{n}}_{2,i,k}$ in different radial bins:   
\begin{equation}
    \mathbf{\bar{n}}_{2,i,k}=\mathcal{A}_{i,k}/w_{{\rm{p}},k}\,\,.
\end{equation}
The $N_{\rm{sub}}$ $\bar{n}_2$ arrays in the same redshift bins are assumed to have the same statistical power, so we can calculate the mean value and covariance matrices for the $\bar{n}_2$ arrays in each redshift:
\begin{equation}
    \mathbf{\bar{n}}_{2,i}=\sum_{k=1}^{N_{\rm{sub}}}\mathbf{\bar{n}}_{2,i,k}/N_{\rm{sub}}\,\,,
\end{equation}
\begin{equation}
    C_{ab,i} = \frac{N_{\rm{sub}}-1}{N_{\rm{sub}}}\sum_{k=1}^{N_{\rm{sub}}}(\bar{n}_{2,i,k}^a-\bar{n}_{2,i}^a)(\bar{n}_{2,i,k}^b-\bar{n}_{2,i}^b)\,\,,
\end{equation}
where $a$ and $b$ denotes the $a$th and $b$th radial bins. We define the $\chi^2$ as:
\begin{equation}
    \chi^2=\sum_{i=1}^{N_{\rm{r}}}(\mathbf{\bar{n}}_{2,i}-\bar{n}_2)^{T}\mathbf{C}^{-1}(\mathbf{\bar{n}}_{2,i}-\bar{n}_2)\,\,,
\end{equation}
where $\bar{n}_2$ is a constant to be determined, $\mathbf{C}^{-1}$ is the inverse of $\mathbf{C}$ and $T$ denotes matrix transposition.

We further split $\rm{pop}_1$ into $N_{\rm{m}}$ mass bins. In principle, $\bar{n}_2$ from different $\rm{pop}_1$ mass bins should be the same. Comparing the fittings in different mass bins can give a consistency check of the data and method, although this may not be done in the smallest mass bins where the numbers of galaxies are too few. So the final $\chi^2_{\rm{all}}$ is defined as:
\begin{equation}
    \chi^2_{\rm{all}}=\sum_{m=1}^{N_{\rm{m}}}\chi^2_{\rm{m}}\,\,,
\end{equation}
where $\chi^2_{\rm{m}}$ is the $\chi^2$ for the $m$th $\rm{pop}_1$ mass bin. We use the Markov chain Monte Carlo (MCMC) sampler {\tt{emcee}} \citep{2013PASP..125..306F} to perform a maximum likelihood analysis of $\{\bar{n}_2\}$.

\begin{figure*}
    \plottwo{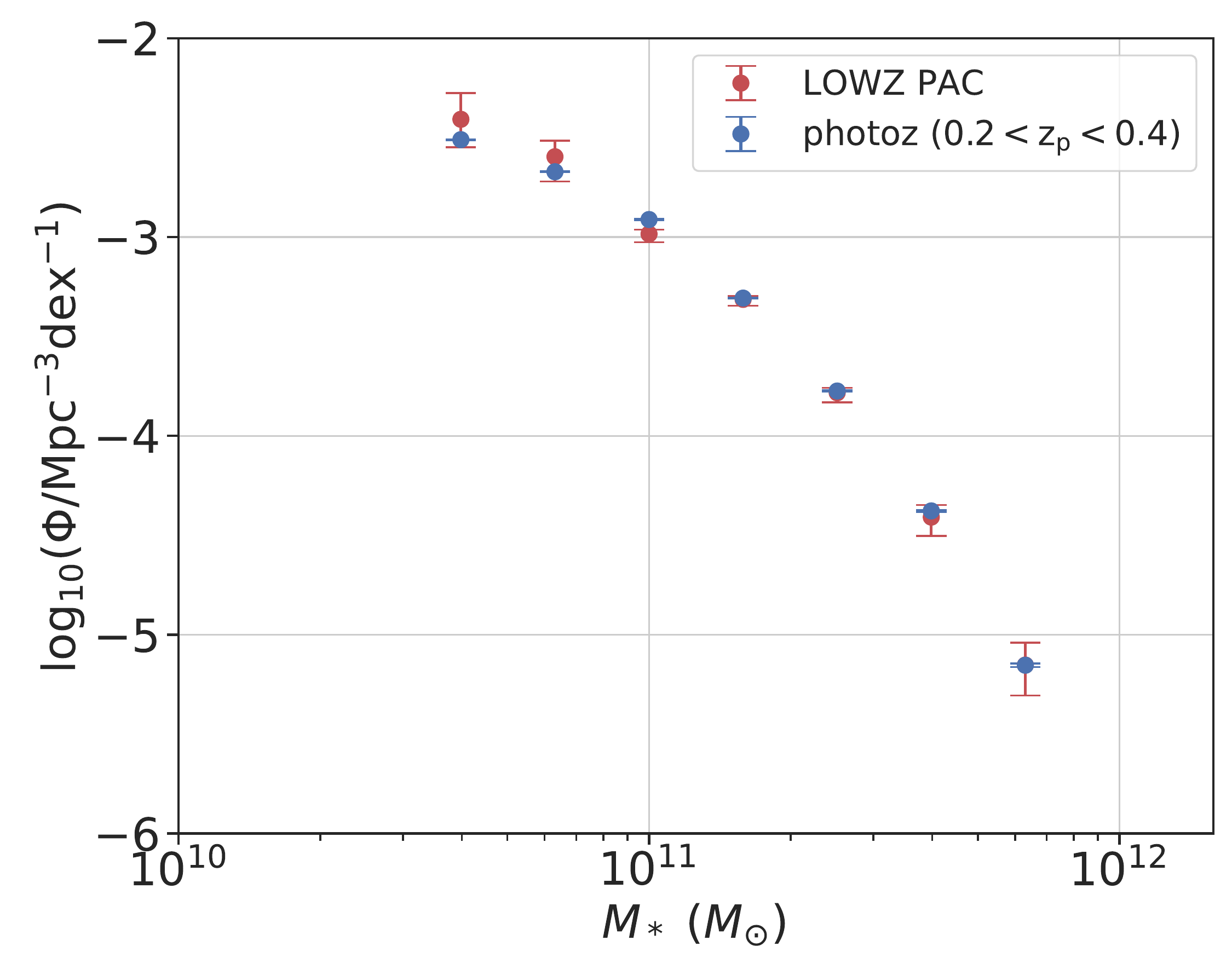}{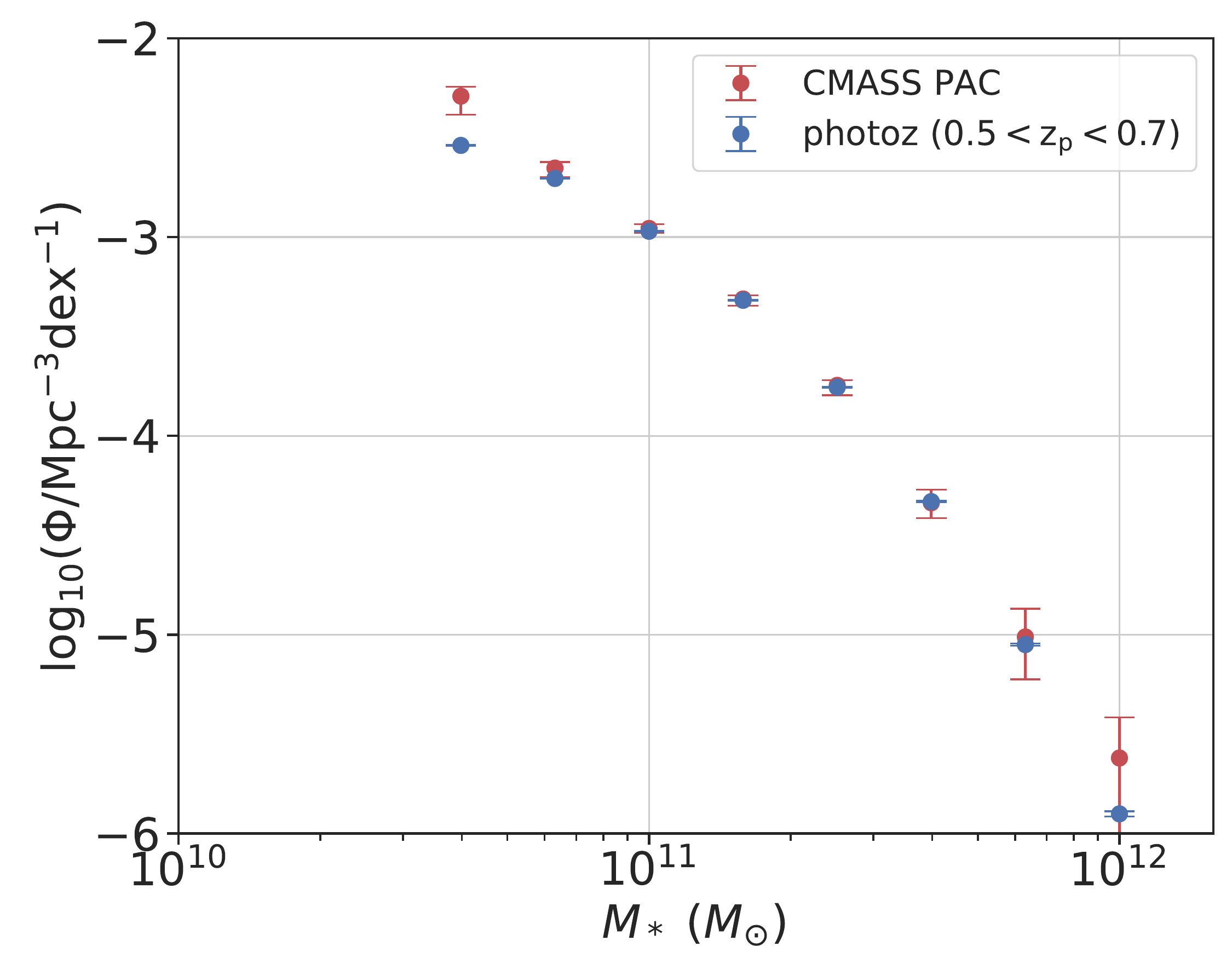}
    \caption{Comparisons of the GSMFs from PAC (red) and DECaLS photo-z (blue; \citealt{2021MNRAS.501.3309Z}) at the massive ends for the LOWZ (left; $0.2<z_p<0.4$) and CMASS (right; $0.5<z_p<0.7$) redshift ranges.}
    \label{fig:fig5}
\end{figure*}

\begin{deluxetable}{ccccc}
\label{tab:t2}
\tablenum{2}
\tablecaption{Parameters of the double Schechter function for the GSMF from the Main sample ($z_s<0.2$).}
\tablewidth{0pt}
\tablehead{
\colhead{$\log_{10}\phi_0$}&\colhead{$\log_{10}M_1$}&\colhead{$\log_{10}M_2$}&\colhead{$\alpha_1$}&\colhead{$\alpha_2$}\\
\colhead{$(\rm{Mpc}^{-3})$}&\colhead{($M_{\odot}$)}&\colhead{($M_{\odot}$)}&\colhead{}&\colhead{}
}
\renewcommand{\arraystretch}{1.5}
\startdata
$-13.58_{-0.31}^{+0.24}$&$10.92_{-0.09}^{+0.11}$&$9.18_{-0.57}^{+0.57}$&$-1.22_{-0.15}^{+0.15}$&$-1.97_{-0.87}^{+0.90}$\\
\enddata
\end{deluxetable}

\section{results}\label{sec:results}
In this section, we present the GSMF measurements for the three redshift ranges. 

\subsection{Galaxy stellar mass functions}
We estimate $\bar{n}_2$ at different stellar masses to get the GSMFs for the three redshift ranges according to the designs in Table \ref{tab:t1}. For every mass bin, $w_{\rm{p}}$ and $\mathcal{A}$ are calculated in the radial range of $0.1h^{-1}{\rm{Mpc}}<r_{{\rm{p}}}<15 h^{-1}\rm{Mpc}$ with $N_{\rm{sub}}=100$. 

Comparisons between $w_{\rm{p}}$ and the modeled $\mathcal{A}/\bar{n}_2$ for different mass bins of $\rm{pop}_1$ and $\rm{pop}_2$ are shown in Figure \ref{fig:figa1}, \ref{fig:figa2} and \ref{fig:figa3} for the Main sample, LOWZ and CMASS, respectively. Results for the same $\rm{pop}_2$
from different $\rm{pop}_1$ masses are consistent with each other, confirming that our methods and results are reliable. Moreover, in the Main sample, $w_{\rm{p}}$ and $\mathcal{A}/\bar{n}_2$ are not vary with the $\rm{pop}_2$ mass at the low mass end ($<10^{10.0}M_{\odot}$), which is in good agreement with the expected nearly constant bias for small galaxies (halos) \citep{1998ApJ...503L...9J,2000MNRAS.318.1144P,2010ApJ...724..878T,2018PhR...733....1D}. 

The GSMFs for the three redshift ranges are shown in Figure \ref{fig:fig4} and are also listed in Table \ref{tab:tb1}. At the high mass end ($>10^{10.6}M_{\odot}$), which covered by three redshifts, the GSMF shows nearly no evolution with redshift. From the low redshift results of the Main sample, our measurements confirm an upturn and a steepening at the low mass end of the GSMF, which has been reported in some previous studies \citep{2006A&A...445...29P,2008MNRAS.388..945B,2010ApJ...724..878T,2012MNRAS.421..621B,2021MNRAS.503.4413M,2022MNRAS.513..439D} . 

To characterise the upturn structure, we fit the GSMF at the low redshift with a double Schechter function with five parameters \citep{2006A&A...445...29P}:
\begin{align}
    \phi(M_{*})dM_{*} = \phi_0\bigg[\left(\frac{M_{*}}{M_1}\right)^{\alpha_1}\exp{\left(-\frac{M_{*}}{M_1}\right)}\\ \notag
    +\left(\frac{M_1}{M_2}\right)\left(\frac{M_{*}}{M_2}\right)^{\alpha_2}\exp{\left(-\frac{M_{*}}{M_2}\right)}\bigg]dM_{*}\,\,,\label{eq:8}
\end{align}
where $\phi(M_*)dM_*$ is the number density of galaxies with mass between $M_*$ and $M_*+dM_*$. We always choose $M_1>M_2$ such that the second term dominates at lower masses. Since the number density $\Phi$ we measured is in an equal logarithmic interval, we have
\begin{equation}
    \Phi(M_*) = \ln{(10)}M_*\phi(M_*)\,\,.
\end{equation}
The best-fit double Schechter function is shown in Figure \ref{fig:fig4} with a black line, with the parameters listed in Table \ref{tab:t2}. 

The galaxy stellar mass density (GSMD) can be obtained by integrating over the GSMF. We report a GSMD of $\rho_*=(2.33\pm0.36)\times10^{8}M_{\odot}\rm{Mpc}^{-3}$ and $\Omega_*=\rho_*/\rho_{\rm{crit}}=(1.67\pm0.25)\times10^{-3}$ at $z_s<0.2$ from our measurements, with an upper limit of $10^{12.5}M_{\odot}$ and a lower limit of $10^{6.0}M_{\odot}$ for the integration.

The GSMF from GAMA DR4 \citep{2022MNRAS.513..439D} at $z_s<0.1$ is also shown in Figure \ref{fig:fig4} for comparison. We find that for $10^9<M_*<10^{10}M_{\odot}$, their measurement agree with ours almost perfectly. However, at the higher mass , their GSMF is higher than ours by $<0.3 dex$, while at the lower mass, our GSMF is higher. The reason for the discrepancies is not fully understood especially for the range of $10^{10.3}<M_*<10^{11.3}M_{\odot}$. For the lowest stellar mass $M_* <10^9 M_{\odot}$, the volume surveyed by GAMA is much smaller and our measurements show a slightly steeper upturn.

\subsection{comparisons with photometric redshift} \label{subsec:4.2}
For a survey like DECaLS, the photometric redshift can be measured accurately at the bright end and at intermediate redshifts ($z_s\sim 0.5$) as shown in many studies \citep{2016A&A...590A.102M,2021MNRAS.501.3309Z,2022arXiv220613633N}, since there are enough spectroscopically identified galaxies covering the whole galaxy population for training, the photometric magnitudes are accurate, and the important spectral features like the $4000 \rm{\AA}$ break are well sampled. 

To verify whether photo-zs from the wide photometric surveys are reliable for the GSMF study, we calculate the stellar mass for the DECaLS galaxies using the photo-z from \citet{2021MNRAS.501.3309Z} estimated using the DECaLS and WISE \citep{2010AJ....140.1868W} photometry. We compare the GSMFs from the PAC and photo-z measurements in Figure \ref{fig:fig5} for the LOWZ ($0.2<z_p<0.4$) and CMASS ($0.5<z_p<0.7$) redshift ranges. Errors of the GSMFs from photo-z are estimated using jackknife re-sampling. We also list the measurements from photo-z in Table \ref{tab:tb1}.

The two independent measurements from PAC and photo-z are in excellent agreement with each other in both redshift ranges at least for $M_{*}>10^{11.0}M_{\odot}$. Although for $M_{*}=10^{12.0}M_{\odot}$ in the CMASS redshift range, the mean value from PAC are higher than that from photo-z, the error of PAC in this mass bin is large. This consistency confirms both that our PAC method for GSMF measurements is reliable and that photo-z from wide photometric surveys estimated from only a few bands is suitable for GSMF studies at intermediate redshifts ($0.2<z_p<0.7$) and massive ends ($M_{*}>10^{11.0}M_{\odot}$). Thus, we verify that photo-z are suitable for LRG target selection and mass completeness studies (\citealt{2016MNRAS.457.4021L,2020RNAAS...4..181Z}; K. Xu et al. 2022, in preparation). 

\section{Conclusion}\label{sec:conclusion}
In this paper, we provide a model independent method for measuring the GSMF by combining the PAC measurements of $\bar{n}_2w_{\rm{p}}$ and spectroscopic measurements of $w_{\rm{p}}$. We apply this method to the photometric catalog from the DESI Legacy Imaging Surveys and the spectroscopic catalogs from the SDSS Main sample ($z_s<0.2$), LOWZ sample ($0.2<z_s<0.4$) and CMASS sample ($0.5<z_s<0.7$), and obtain the GSMFs at the three redshifts down to $10^{8.2}M_{\odot}$, $10^{10.6}M_{\odot}$ and $10^{10.6}M_{\odot}$, respectively. 

At the high mass end ($>10^{10.6}M_{\odot}$), our measurement shows that there is no evolution of GSMF since $z_s=0.6$. At the low mass end ($<10^{9.0}M_{\odot}$), we find an upturn in the GSMF at $z_s<0.2$, which is slightly steeper than reported in \citet{2022MNRAS.513..439D} using the GAMA DR4 data. We also report a galaxy stellar mass density (GSMD) of $\rho_*=(2.33\pm0.36)\times10^{8}M_{\odot}\rm{Mpc}^{-3}$ and $\Omega_*=(1.67\pm0.25)\times10^{-3}$ at $z_s<0.2$.  

We compare the PAC measurements of the GSMFs with the DECaLS photo-z measurements and find that the two results are in excellent agreement with each other at intermediate redshifts ($0.2<z_p<0.7$) and massive ends ($M_{*}>10^{11.0}M_{\odot}$), verifying both that our method is reliable and photo-z is also suitable for LRG target selection and mass completeness studies. The accurate measurements of the GSMFs for $z_s<0.7$ also provide a testbed for semi-analytical models and hydrodynamical simulations of galaxy formation.

Our result also shows that our method can achieve a relatively good measurement of GSMF with a spectroscopic sample that is even highly incomplete ($\sim 10^{-3}$) in stellar mass (cf the results for the LOWZ and CMASS samples). With our method, spectroscopic surveys with low target sampling rates can also be useful for GSMF studies.

With the next generation large and deep photometric and spectroscopic surveys such as Dark Energy Spectroscopic Instrument (DESI, \citealt{2016arXiv161100036D}), Legacy Survey of Space and Time (LSST, \citealt{2019ApJ...873..111I}) and Euclid \citep{2011arXiv1110.3193L}, we can extend the GSMF measurements to higher redshift and smaller stellar mass.

\begin{acknowledgments}
The work is supported by NSFC (12133006, 11890691, 11621303) and by 111 project No. B20019. We gratefully acknowledge the support of the Key Laboratory for Particle Physics, Astrophysics and Cosmology, Ministry of Education. This work made use of the Gravity Supercomputer at the Department of Astronomy, Shanghai Jiao Tong University.

This publication has made use of data products from the Sloan Digital Sky Survey (SDSS). Funding for SDSS and SDSS-II has been provided by the Alfred P. Sloan Foundation, the Participating Institutions, the National Science Foundation, the U.S. Department of Energy, the National Aeronautics and Space Administration, the Japanese Monbukagakusho, the Max Planck Society, and the Higher Education Funding Council for England.

Funding for SDSS-III has been provided by the Alfred P. Sloan Foundation, the Participating Institutions, the National Science Foundation, and the U.S. Department of Energy Office of Science. The SDSS-III web site is http://www.sdss3.org/. SDSS-III is managed by the Astrophysical Research Consortium for the Participating Institutions of the SDSS-III Collaboration including the University of Arizona, the Brazilian Participation Group, Brookhaven National Laboratory, Carnegie Mellon University, University of Florida, the French Participation Group, the German Participation Group, Harvard University, the Instituto de Astrofisica de Canarias, the Michigan State/Notre Dame/JINA Participation Group, Johns Hopkins University, Lawrence Berkeley National Laboratory, Max Planck Institute for Astrophysics, Max Planck Institute for Extraterrestrial Physics, New Mexico State University, New York University, Ohio State University, Pennsylvania State University, University of Portsmouth, Princeton University, the Spanish Participation Group, University of Tokyo, University of Utah, Vanderbilt University, University of Virginia, University of Washington, and Yale University.

The Legacy Surveys consist of three individual and complementary projects: the Dark Energy Camera Legacy Survey (DECaLS; Proposal ID \#2014B-0404; PIs: David Schlegel and Arjun Dey), the Beijing-Arizona Sky Survey (BASS; NOAO Prop. ID \#2015A-0801; PIs: Zhou Xu and Xiaohui Fan), and the Mayall z-band Legacy Survey (MzLS; Prop. ID \#2016A-0453; PI: Arjun Dey). DECaLS, BASS and MzLS together include data obtained, respectively, at the Blanco telescope, Cerro Tololo Inter-American Observatory, NSF’s NOIRLab; the Bok telescope, Steward Observatory, University of Arizona; and the Mayall telescope, Kitt Peak National Observatory, NOIRLab. The Legacy Surveys project is honored to be permitted to conduct astronomical research on Iolkam Du’ag (Kitt Peak), a mountain with particular significance to the Tohono O’odham Nation.
\end{acknowledgments}

\newpage

\bibliography{sample631}{}
\bibliographystyle{aasjournal}


\appendix
\restartappendixnumbering
\section{Measurements and  fittings}\label{sec:A}
\begin{figure}[h]
    \plotone{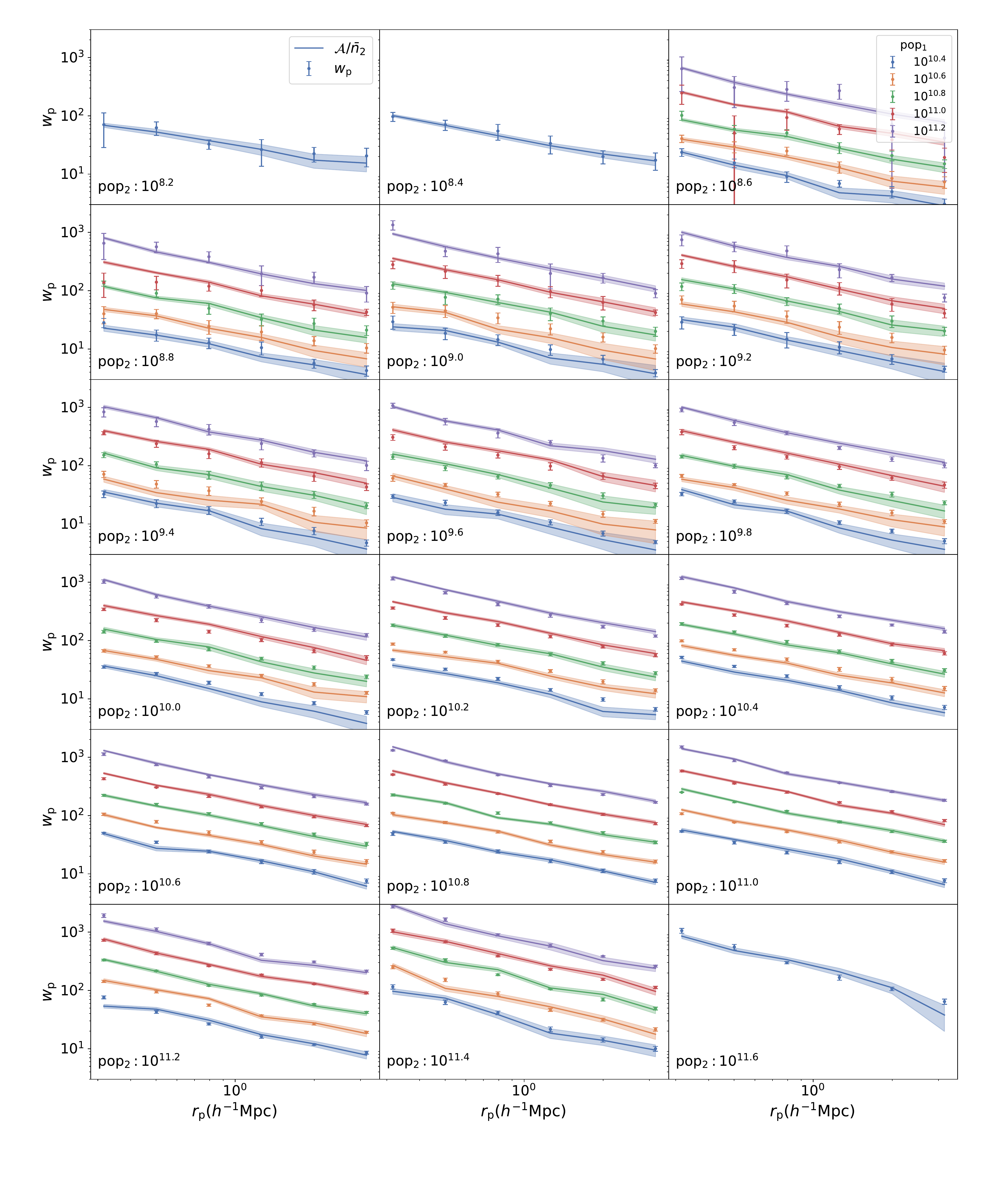}
    \caption{Comparing $\mathcal{A}/\bar{n}_2$ (lines) with $w_{\rm{p}}$ (dots) for the Main sample. Each panel is for the same mass bin of $\rm{pop}_2$. Different colors are for different mass bins of $\rm{pop}_1$ and results are multiplied by $0.25$, $0.5$, $1$, $2$ and $4$ for better illustration.}
    \label{fig:figa1}
\end{figure}

\begin{figure}[h]
    \plotone{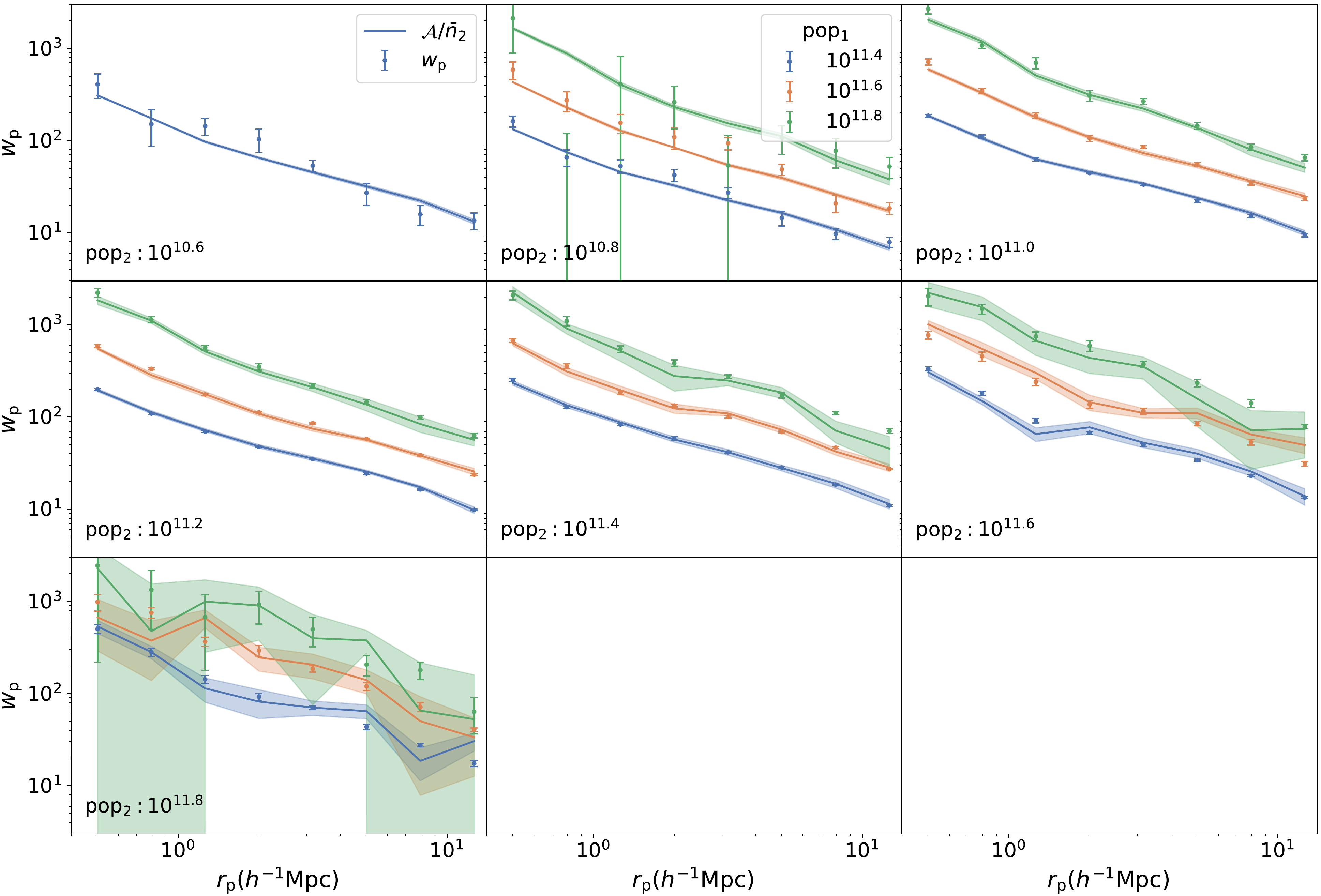}
    \caption{The same as Figure \ref{fig:figa1} but for LOWZ.}
    \label{fig:figa2}
\end{figure}

\begin{figure}[h]
    \plotone{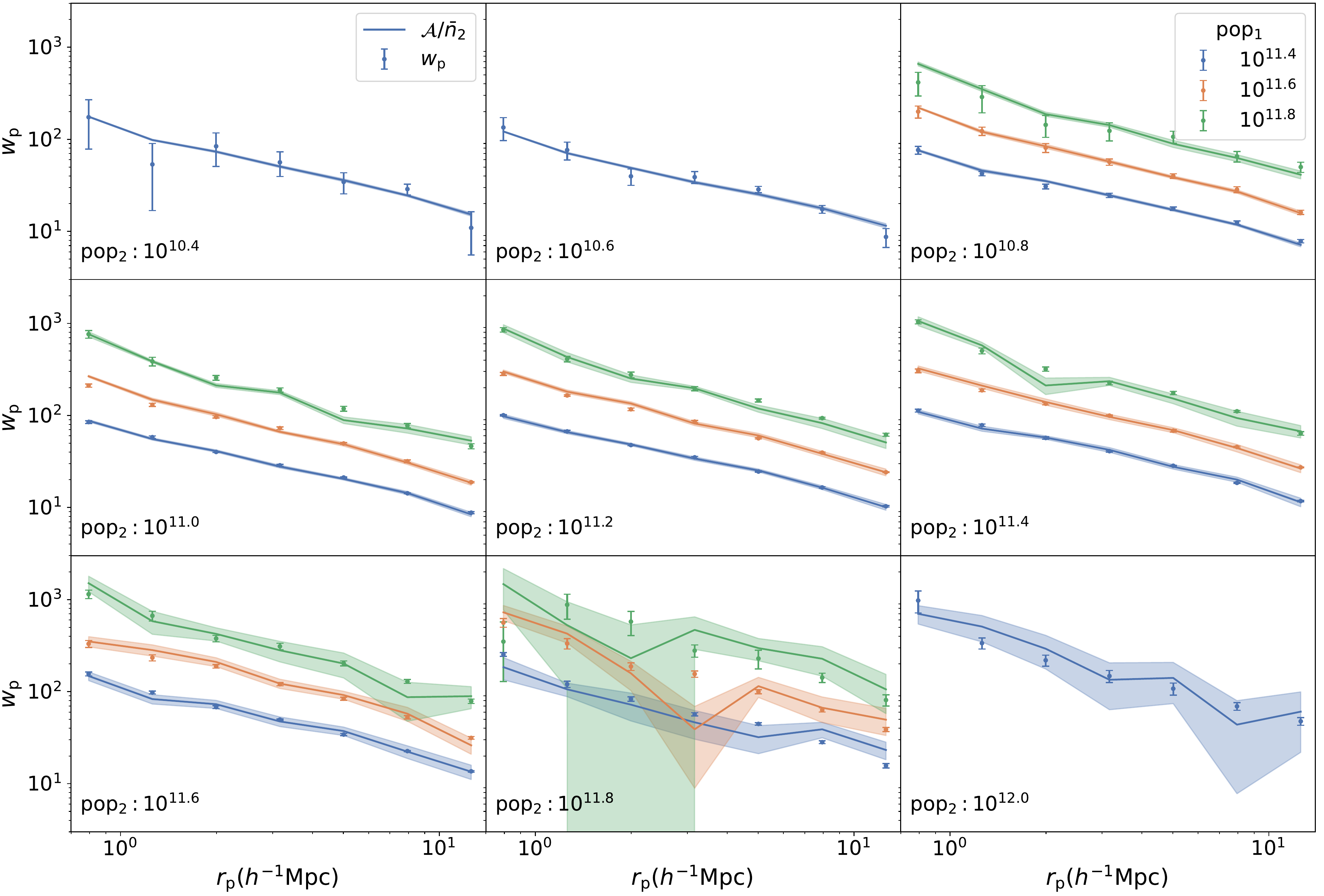}
    \caption{The same as Figure \ref{fig:figa1} but for CMASS.}
    \label{fig:figa3}
\end{figure}

In Figure \ref{fig:figa1}, \ref{fig:figa2} and \ref{fig:figa3}, we check the modeling results of $\bar{n}_2$ by comparing $\mathcal{A}/\bar{n}_2$ with $w_{\rm{p}}$ for the Main sample, LOWZ and CMASS, respectively. Dots with error bars show the results for $w_{\rm{p}}$ and lines with shadows are the results for $\mathcal{A}/\bar{n}_2$. Each panel show the results for the same mass bin of $\rm{pop}_2$ and different colors are for different mass bins of $\rm{pop}_1$. In all mass bins, $\mathcal{A}/\bar{n}_2$ is in good agreement with $w_{\rm{p}}$, confirming that our methods and results are robust and reliable. 

\section{The GSMFs in Tabular form}\label{sec:B}
\begin{table*}[h]
    \centering
    \caption{The galaxy stellar mass number-density distributions $\log_{10}(\Phi/{\rm{Mpc^{-3}dex^{-1})}}$ at different redshifts from PAC ($z_s$) and DECaLS photo-z ($z_p$).}
    \begin{threeparttable}
    \begin{tabular}{cccccc}
     \toprule
     $\log_{10}(M_{*}/M_{\odot})$ & $z_s<0.2$ & $0.2<z_s<0.4$ & $0.5<z_s<0.7$ & $0.2<z_p<0.4$ & $0.5<z_p<0.7$\\
     \midrule
     8.2 & $-1.36_{-0.22}^{+0.20}$ & &\\
     8.4 & $-1.35_{-0.15}^{+0.11}$ & &\\
     8.6 & $-1.47_{-0.16}^{+0.12}$ & &\\
     8.8 & $-1.62_{-0.10}^{+0.11}$ & &\\
     9.0 & $-1.76_{-0.15}^{+0.11}$ & &\\
     9.2 & $-1.90_{-0.15}^{+0.14}$ & &\\
     9.4 & $-1.99_{-0.14}^{+0.08}$ & &\\
     9.6 & $-2.08_{-0.11}^{+0.07}$ & &\\
     9.8 & $-2.08_{-0.08}^{+0.06}$ & &\\
     10.0 & $-2.14_{-0.08}^{+0.04}$ & &\\
     10.2 & $-2.23_{-0.05}^{+0.04}$ & &\\
     10.4 & $-2.30_{-0.05}^{+0.04}$ & & \\
     10.6 & $-2.43_{-0.04}^{+0.04}$ & $-2.41_{-0.14}^{+0.13}$ & $-2.29_{-0.09}^{+0.05}$ & $-2.505_{-0.002}^{+0.002}$ & $-2.538_{-0.001}^{+0.001}$\\
     10.8 & $-2.60_{-0.03}^{+0.03}$ & $-2.60_{-0.12}^{+0.08}$ & $-2.65_{-0.05}^{+0.03}$ & $-2.667_{-0.002}^{+0.002}$ & $-2.705_{-0.001}^{+0.001}$\\
     11.0 & $-2.85_{-0.03}^{+0.03}$ & $-2.98_{-0.04}^{+0.02}$ & $-2.96_{-0.02}^{+0.02}$ & $-2.905_{-0.002}^{+0.002}$ & $-2.970_{-0.001}^{+0.001}$\\
     11.2 & $-3.20_{-0.05}^{+0.03}$ & $-3.31_{-0.03}^{+0.02}$ & $-3.31_{-0.03}^{+0.02}$ & $-3.298_{-0.003}^{+0.003}$ & $-3.318_{-0.002}^{+0.002}$\\
     11.4 & $-3.80_{-0.08}^{+0.04}$ & $-3.78_{-0.05}^{+0.02}$ & $-3.75_{-0.05}^{+0.03}$ & $-3.764_{-0.004}^{+0.004}$ & $-3.756_{-0.002}^{+0.002}$\\
     11.6 & $-4.49_{-0.12}^{+0.05}$ & $-4.41_{-0.09}^{+0.06}$ & $-4.34_{-0.08}^{+0.07}$ & $-4.366_{-0.005}^{+0.005}$ & $-4.330_{-0.004}^{+0.004}$\\
     11.8 & & $-5.15_{-0.15}^{+0.11}$ & $-5.01_{-0.21}^{+0.14}$ & $-5.141_{-0.009}^{+0.009}$ & $-5.050_{-0.006}^{+0.006}$\\
     12.0 & & & $-5.61_{-0.40}^{+0.20}$ & & $-5.901_{-0.014}^{+0.013}$\\
     \bottomrule
    \end{tabular}  
    \end{threeparttable}
    \label{tab:tb1}
\end{table*}
In Table \ref{tab:tb1}, we list the mean values and errors of the estimated GSMFs $\log_{10}(\Phi/{\rm{Mpc^{-3}dex^{-1})}}$ at different redshift ranges from both PAC ($z_s$) and DECaLS photo-z ($z_p$).

\section{Systematics between DECaLS and BASS+MzLS}\label{sec:C}

\begin{figure}[th]
    \plotone{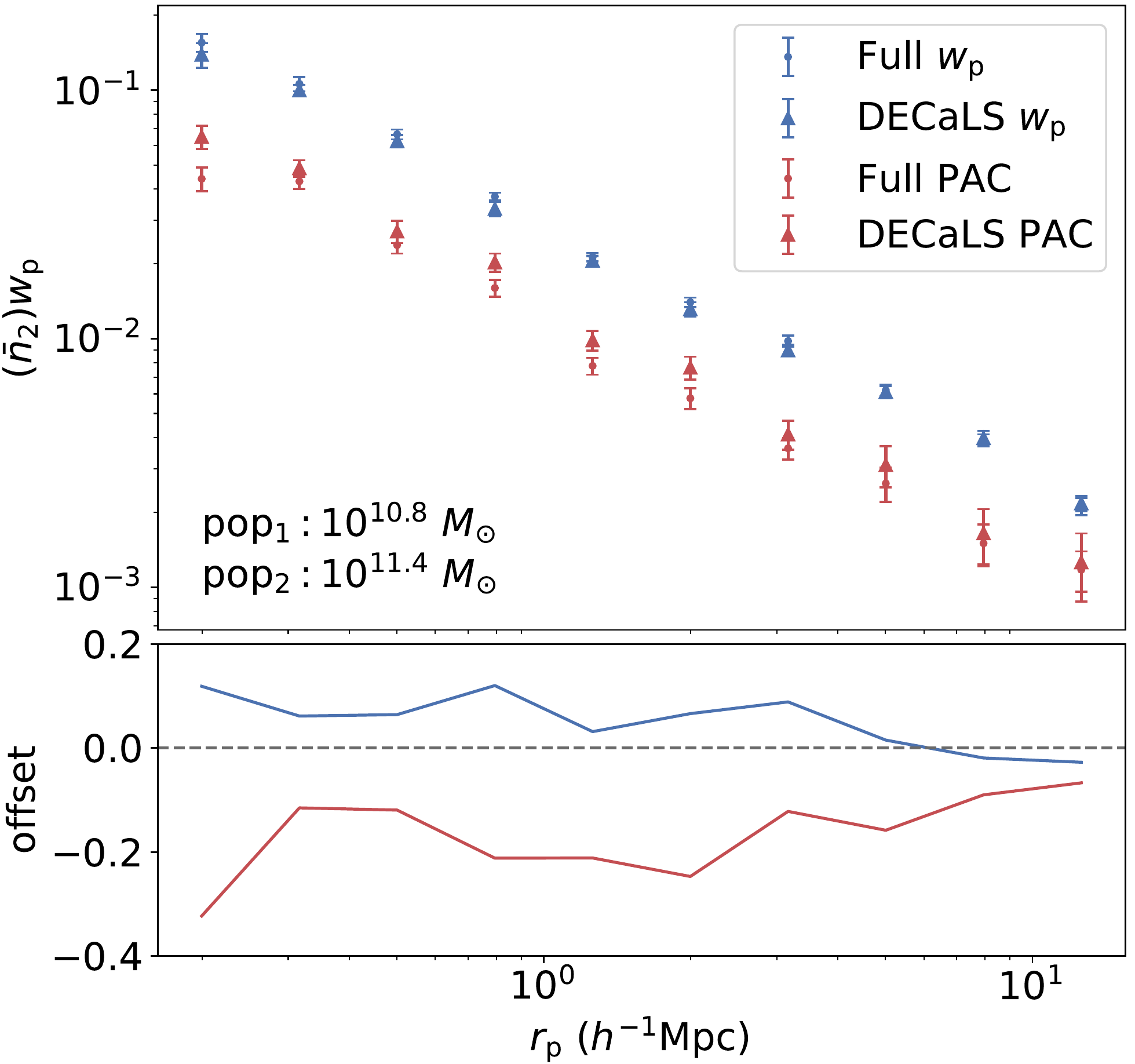}
    \caption{Comparing $w_{\rm{p}}$ (blue) and PAC $\bar{n}_2w_{\rm{p}}$ (red) measurements of the Full (DECaLS+BASS+MzLS) samples (dots) and DECaLS only samples (triangles). Measurements are shown in the top panel and the relative offsets of the Full samples from the DECaLS samples are shown in the bottom panel. The comparisons are for $\rm{pop}_1$ with $10^{10.8}M_{\odot}$ and $\rm{pop}_2$ with $10^{11.4}M_{\odot}$ at $z_s<0.2$. $w_{\rm{p}}$ is divided by $5000$ for better comparison.}
    \label{fig:fig1}
\end{figure}

In Figure \ref{fig:fig1}, we compare the $w_p$ and $\bar{n}_2w_p$ measurements from the DECaLS samples alone and the full DECaLS+BASS+MzLS samples, with ${\rm{pop}}_1$ of $10^{10.8}M_{\odot}$ and ${\rm{pop}}_2$ of $10^{11.4}M_{\odot}$ at $z_s<0.2$. As shown in the figure, there are both systematics in the measurements of $w_p$ and $\bar{n}_2w_p$ for the two samples, but the relative offset is much smaller in $w_p(<10\%)$ than in $\bar{n}_2w_p(>20\%)$, which may due to the much faster changes of GSMF with stellar mass than galaxy bias at the high mass ends. Thus, we decide to use both DECaLS and BASS+MzLS for the $w_p$ measurements while use only DECaLS for the PAC measurements. In this way, we can reduce the statistical uncertainties in the measurements of $w_p$ and avoid the large systematics from the measurements $\bar{n}_2w_p$. 
\end{document}